\documentclass[]{spie}  
\newcommand{\GO}[1]{{{\textcolor{black}{#1}}}}
 
\usepackage{amsmath,amsfonts,amssymb}
\usepackage{graphicx}
\usepackage[colorlinks=true, allcolors=blue]{hyperref}
\usepackage{caption}
\usepackage{subcaption}
\newcommand{\SP}[1]{{{\textcolor{black}{#1}}}}

\title{Quantum reading: the experimental set-up} 

\author[b]{Elena Losero}
\author[a]{Giuseppe Ortolano}
\author[a]{Fabio Saccomandi}
\author[a]{Ivano Ruo-Berchera}
\author[c]{Stefano Pirandola}
\author[a]{Marco Genovese}
\affil[a]{Quantum metrology and nano technologies division,  INRiM,  Strada delle Cacce 91, 10135 Torino, Italy}
\affil[b]{Laboratory of Quantum Nano-Optics, EPFL, 1015, Lausanne, Switzerland}
\affil[c]{Department of Computer Science, University of York, York YO10 5GH, United Kingdom}

\authorinfo{Further author information: Elena Losero: E-mail: elena.losero@epfl.ch \\ Giuseppe Ortolano: E-mail: giuseppe.ortolano@polito.it}

\begin{document} 
\maketitle

\begin{abstract}
The protocol of quantum reading refers to the quantum enhanced retrieval of information from an optical memory, whose generic cell stores a bit of information in two possible lossy channels. In the following we analyze the case of a particular class of optical receiver, based on photon counting measurement, since they can be particularly simple in view of real applications. We show that a quantum advantage \SP{is achievable when a transmitter based on two-mode squeezed vacuum (TMSV) states is combined with a photon counting receiver, and we experimentally confirm it}. 
In this paper, after introducing some theoretical background, we focus on the experimental realisation, describing the data collection and the data analysis in detail.
\end{abstract}


\section{INTRODUCTION}
\SP{A general strategy \GO{to read} data} a classical digital memory can be modeled as follows \cite{pirandola11}: a bipartite bosonic system, called transmitter $T(\rho,M,L)$, in a state $\rho$ formed by a signal system with $M$ modes and an idler system with $L$ modes, is used to discriminate between two values of transmittance $\tau_{\mu}$ storing the value
of a classical bit, $u = {0, 1}$. The signal system probes a single cell of the memory while the
idler system goes directly to the receiver $R$, where an appropriate joint measurement of the final
bipartite state is performed to retrieve the value of the bit with a certain probability of error $P_{err}$. Given this scenario the goal is to minimize $P_{err}$ under the energy \SP{constraint} where $N$ photons in total are addressed to the memory cell. Assuming an optimal measurement at the receiver, the minimization of $P_{err}$ over all the possible transmitters with fixed signal energy is difficult to solve.
However, in the relevant case of a classical state $\rho$, having a positive Glauber–Sudarshan P representation, a \GO{lower bound} on the error probability can be found. In particular in \GO{Ref.} \cite{pirandola11} the error probability for discrimination between the final states $\mathcal{E}_0(\rho)$ and $\mathcal{E}_1(\rho)$, where $\mathcal{E}_{\mu}$ represents the action of the attenuation channel induced by $\tau_{\mu}$, is found to be: 

\begin{equation} \label{cl}
P^{cla}_{err}\geq\mathcal{C}(N,\tau_0,\tau_1)=\frac{1-\sqrt{1-e^{-N(\sqrt{\tau_1}-\sqrt{\tau_0})^2}}}{2}.
\end{equation}

Here \GO{${C}(N,\tau_0,\tau_1)$ sets an absolute limit on the performance that can be achieved using any transmitter in a classical state}. Note that \GO{it} depends only on the
mean number of signaling photons $N$ and the values of the transmittance $\tau_0$ and $\tau_1$. Note also that this bound is not necessarily tight, i.e. it is not known if it exists a classical measurement scheme able to saturate it.
A probability of error $P_{err}$ means that on average  $1-H(P_{err})$ bits are recovered, where $H(\cdot)$ denotes the binary Shannon entropy \cite{cover96}. It can be formally proven \cite{pirandola11} that this limit can be beaten by a certain class of quantum transmitters, namely EPR correlated ones, \GO{assuming an unspecified optimal measurement at the receiver}. \GO{Denoting the probability of error of a quantum transmitter as  $P_{err}^{qua}$}, \GO{we can evaluate the gain in bits over the classical bound as:}
\begin{equation}\label{eq:gain}
G_a(N,\tau_0,\tau_1)=1-H(P_{err}^{qua})-(1-H(\mathcal{C}(N,\tau_0,\tau_1))).
\end{equation}
\section{PHOTON COUNTING DISCRIMINATION STRATEGIES}
The macroscopic effect of an attenuation channel (e.g. $\mathcal{E}_0$ and $\mathcal{E}_1$) is a reduction of the main energy of the system. In particular, considering an electromagnetic field, the mean number of photons $N$ after a channel characterized by a transmission $0\leq \tau\leq1$, will be reduced to $\tau N$. From this consideration, a natural approach \GO{to discriminate the channels is} to \GO{perform a} direct photon counting measurement. In the case of parameter estimation, when the goal is to estimate the continuous parameter $\tau$, it can be proven that a photon counting measurement combined with Fock states or high number of TMSV replicas is actually able to reach the \SP{ultimate quantum limit.} 
For the discrete case, i.e. the discrimination problem considered here, such a \SP{proof} has not been given before our work. Note that this class of discrimination strategies is extremely relevant from the experimental point of view since they can be easily implemented.

\subsection{Effect of the attenuation on the photon number distribution}
 
The output of a photon counting measurement on a field $\hat{a}$ in a generic state $\rho$, is a classical random variable $n$ distributed as $P_0(n)=\langle n| \rho | n\rangle$, where $| n\rangle$ is the eigenstate with eigenvalue $n$ of the number operator $\hat{n}=\hat{a}^{\dagger} \hat{a}$ of the field. 
The effect of an attenuation channel $\mathcal{E}_\tau$ on a field can be modeled using a Beam Splitter (BS) with transmission $\tau$ for which the input-output relations are well known: $\hat{a}_t=\sqrt{\tau}\hat{a}+i\sqrt{(1-\tau)}\hat{v}$ and $\hat{a}_r=i\sqrt{(1-\tau)}\hat{a}+\sqrt{\tau}\hat{v}$. Here $\hat{a}$ denotes the input field, $\hat{a}_t$ and $\hat{a}_r$ the transmitted and reflected fields, $\hat{v}$ is the field at the second port of the BS. Thermal noise in the channel can be modeled taking $\hat{v}$ in a thermal state. In the following however thermal noise will be neglected so that $\hat{v}$ will be considered to be in a vacuum state $|0 \rangle$. 

The photon number distribution for the output state is found to be:
\begin{equation}
\GO{\langle N|\mathcal{E}_{\tau}(\rho)|N\rangle = \sum^\infty_{n=N} \langle n|\rho|n\rangle B(n|m,\tau), \label{di}}
\end{equation}
where \GO{$B(n|m,\tau)=\genfrac(){0pt}{0}{n}{N} \tau^{N}(1-\tau)^{n-N}$} is a binomial distribution with $n$ trials and probability of success $\tau$. 
The process can be seen as each photon undergoing a \SP{Bernoulli} trial with probability of success $\tau$ \SP{leading to} a binomial distribution with $n$ trials, $B(n,\tau)$. An arbitrary initial distribution $P_0(n)$ will then be compounded, by $\mathcal{E}_\tau$,  with a binomial distribution $B(n,\tau)$ in accordance with Eq.(\ref{di}).

From here thereafter the quantity considered in the characterization of a state will be its photon number distribution, in terms of which Eq.(\ref{di}) can be rewritten as:
\begin{equation}
P (n)=\sum_{m=0}^\infty P_0(m)B(n|m,\tau)
\end{equation}
In the case of multivariate distributions, the case of interest when considering a multipartite system, the relevant quantity is the joint photon number distribution, $P(n_1,..,n_{\mathcal{N}})$.
Considering an initial distribution $P_0(n_1,..,n_{\mathcal{N}})$ going through $\mathcal{N}$ independent channels with attenuation $\tau=\{\tau_1,...,\tau_\mathcal{N}\}$ the final distribution will be, similarly to the univariate case:
\begin{equation}
P(n_S,..,n_{\mathcal{N}})=\sum_{m_1,...,m_\mathcal{N}=0}^\infty P_0(m_1,..,m_{\mathcal{N}})\prod_i^\mathcal{N} B(n_i|m_i,\tau_i). \label{pt}
\end{equation}

\subsection{Bayesian discrimination}\label{bd}
Suppose the pair of counts $\textbf{n}=(n_S,n_I)$ is the result of a measurement, from the signal and idler channels respectively, after the signal from a known source has passed trough a channel $\mathcal{E}_i$. If we consider only two possible channels, $\mathcal{E}_0$ and $\mathcal{E}_1$ (corresponding to a transmittance $\tau_0$ and $\tau_1$ respectively), a discrimination between the two hypothesis can be done considering the Bayesian  probability $P(\tau_i|\textbf{n})$ given by:
\begin{align}
P(\tau_i|\textbf{n})&=\frac{P(\textbf{n}|\tau_i)P(\tau_i)}{P(\textbf{n})}= \nonumber \\
&=\frac{P(\textbf{n}|\tau_i)}{P(\textbf{n}|\tau_0)+P(\textbf{n}|\tau_1)},
\end{align}
where we are considering the discrimination between two equi-probable channels, i.e. $P(\tau_i)=1/2$.  Given the constant prior, the Bayesian probability is proportional to the likelihood $P(\textbf{n}|\tau_i)$.  

After measurement, the transmittance $\tau_i$ can be chosen such that $P(\tau_i|\textbf{n})\geq P(\tau_{j}|\textbf{n})$, that given the proportionality implies an analogous condition on the likelihood, with probability of success $P(\tau_i|\textbf{n})$. The probability of success in function of the pair $\textbf{n}$ can then be written as $P_s(\textbf{n})=\max_{i=0,1}P(\tau_i|\textbf{n})$. Averaging $P_s(\textbf{n})$ over the pair distribution $P(\textbf{n})=\frac{1}{2} \big( P(\textbf{n}|\tau_0)+P(\textbf{n}|\tau_1) \big)$ yields the expected probability of success,$P_s(\tau_0,\tau_1)$, of the recovery:
\begin{align}\label{ps}
P_s(\tau_0,\tau_1)&=\frac{1}{2}\sum_{\textbf{n}} \max_{i=0,1}P(\tau_i|\textbf{n}) (P(\textbf{n}|\tau_0)+P(\textbf{n}|\tau_1) ) = \nonumber \\
&= \frac{1}{2}\sum_{\textbf{n}} \max_{i=0,1}P(\textbf{n}|\tau_i).
\end{align}  

\subsection{Photon-counting strategy for classical states}\label{sec:cl}
In this section we focus on the explicit discrimination strategy based on photon counting presented in Sec. \ref{bd}, while considering classical states. In particular, we focus on the relevant case of a transmitter in a coherent case $\rho_{coh}$, e.g. the state that can be easily obtained whit a laser. A coherent state presents a Poissonian photon number distribution: $P_0(n)=P_\lambda(n)$. An important property of $P_\lambda(n)$ is that the attenuation process does not change the form of the distribution, i.e. the distribution at the output of the attenuation channel $\mathcal{E}_{\tau}$ is $P_{\lambda\tau}(n)$.

In the following we will consider a single mode coherent state signaling $N$ photons, $T(\rho_{coh},1,0)$. In the coherent case, given the independence of the events described by a Poisson distribution, considering $M$ independent modes rather than a single one would not change the results given that the $N$ photons in the single mode are already independent from each other.

Considering two channels of transmittance $\tau_0<\tau_1$ and a source with initial distribution $P_{\lambda}(n)$, after the channel the distribution will be either $P_{\lambda\tau_0}(n)$ or $P_{\lambda\tau_1}(n)$. As described in section \ref{bd} we will chose $\tau_0$ for each value of $\bar{n}$ such that $P_{\lambda\tau_0}(\bar{n})\geq P_{\lambda\tau_1}(\bar{n})$. 

Solving this condition for $\bar{n}$ yields the threshold:
\begin{equation}\label{eq:thr1}
\bar{n}\leq \frac{\lambda(\tau_1-\tau_0)}{\log(\tau_1/\tau_0)}\doteq n_{th}.
\end{equation}

\begin{figure}[ht]
\centering
\includegraphics[width=0.58\textwidth]{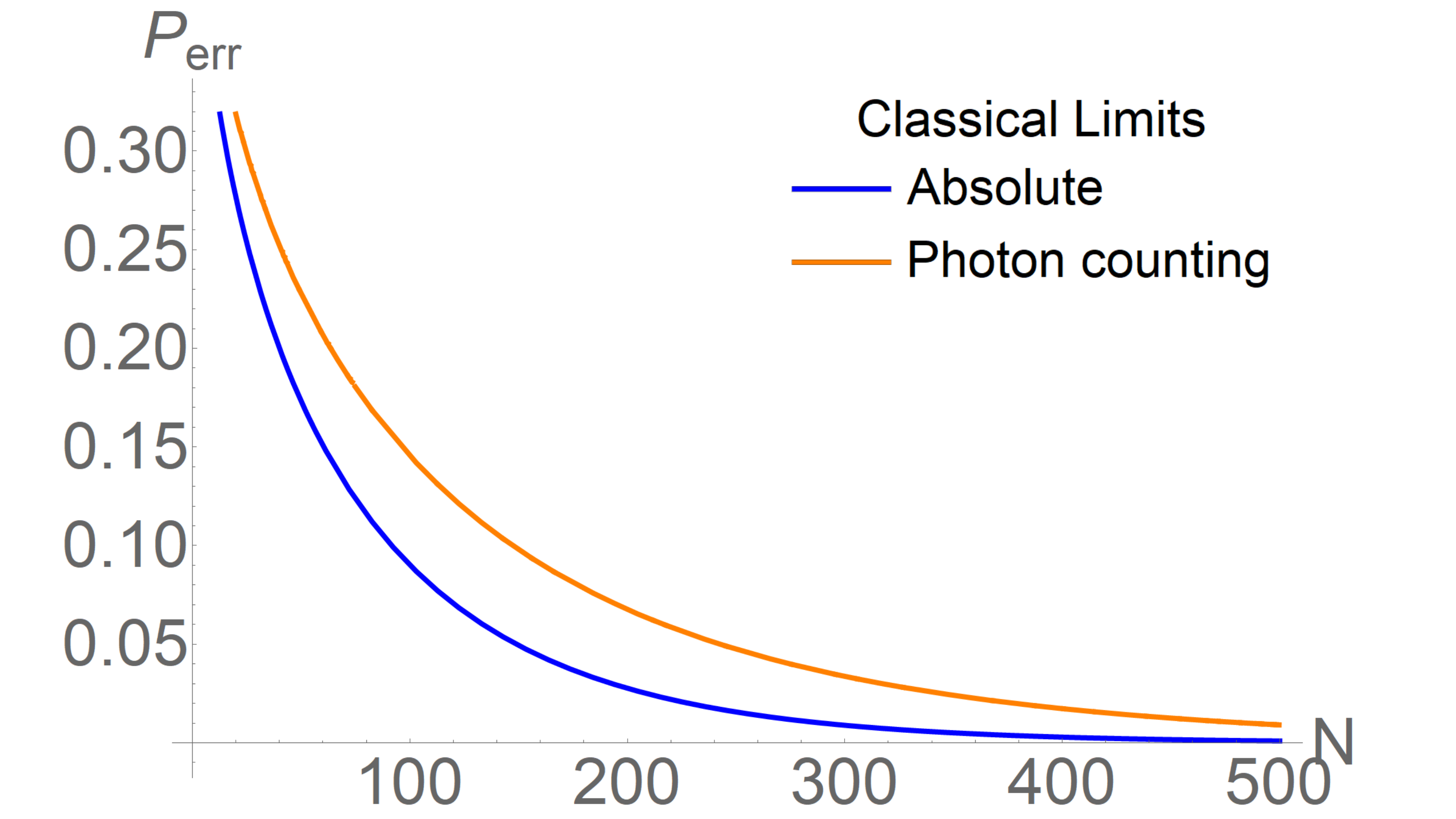} 
\vspace{+0.5cm}
\caption{\label{c} \emph{Comparison of classical limits}. A comparison between the theoretical classical limit, supposing an optimal measurement, reported in Eq.~(\ref{cl}), and the limit for classical states when an explicit photon counting measurement is considered, reported in Eq.~(\ref{eq:coh}).}
\end{figure}

The average probability of success in this case, $P_s^{cla,pc}(\tau_0,\tau_1)$, will then be given by:
\begin{align}
P_s^{cla,pc}(\tau_0,\tau_1)&=\sum_{n=0}^\infty P_s(n)\frac{1}{2}\big( P_{\lambda\tau_0}(n)+ P_{\lambda\tau_1}(n)\big)= \nonumber \\
&=\frac{1}{2}\sum_{n=0}^{n_{th}} P_{\lambda\tau_0}(n) +\frac{1}{2}\sum_{n=n_{th}}^\infty P_{\lambda\tau_1}(n) \nonumber \\
&=\frac{1}{2}\Big( 1+\frac{\Gamma(\lfloor n_{th}+1 \rfloor,\lambda\tau_0)-\Gamma(\lfloor n_{th}+1 \rfloor,\lambda\tau_1)}{\lfloor n_{th} \rfloor !}\Big),
\end{align} 
where $\lfloor x \rfloor$ denotes the floor and $\Gamma(x,y)$ is the incomplete gamma function. The probability of error for the photon counting strategy in the case of a classical input state (i.e. a coherent state) can be computed as:
\begin{equation}\label{eq:coh} P_{err}^{cla,pc}(\tau_0,\tau_1)=1-P_s^{cla,pc}(\tau_0,\tau_1).
\end{equation}
This quantity can be compared with the theoretical classical limit in Eq. (\ref{cl}). A comparison for $\tau_0=0.8$ and $\tau_1=1$ is shown in Fig. \ref{c}. The gain in terms of bits of a quantum strategy having error probability $P^{qua}_{err}$ over the classical photon counting bound is:
\begin{equation}\label{eq:gemp}
    G_{emp}(N,\tau_0,\tau_1)=1-H(P_{err}^{qua})-(1-H(P_{err}^{cla,pc}(\tau_0,\tau_1))).
\end{equation}

\subsection{Photon-counting strategy for a two-mode squeezed vacuum state}\label{sec:twb}
A quantum state, commonly produced in optical laboratories, having correlations that surpass the classical limit is the two-mode squeezed vacuum (TMSV) state, $|\psi\rangle_{\text{TMSV}}$:
\begin{equation}
|\psi\rangle_{\text{TMSV}}=\sum_{n=0}^\infty c_N(n)|n\rangle_S|n\rangle_I,
\end{equation} 
where $S$ and $I$ denote a signal and idler system and $|c_N(n)|^2=\frac{N^n}{(1+N)^{n+1}}$ is a thermal distribution with mean number of photons $N$. The distribution of the count pair $\textbf{n}=(n_S,n_I)$ is then:
\begin{equation}
P(\textbf{n})=|\langle n_S,n_I|\psi\rangle_{\text{TMSV}}|^2= |c_N(n)|^2 \delta(n-n_S)\delta(n-n_I). \label{id}
\end{equation} 
After the signal system passes through a channel $\tau$ this distribution is mapped according to Eq. (\ref{pt}) into $P(\textbf{n}|\tau)$:
\begin{equation}
P(\textbf{n}|\tau)=|c_N(n_I)|^2 B(n_S|n_I,\tau) \label{twbt}.
\end{equation} 
Once again, when discriminating between two channels of transmittance $\tau_0<\tau_1$, the value chosen is the one for which the likelihood in Eq. (\ref{twbt}) is higher. Solving this condition for $n_S$ we get:
\begin{equation}\label{eq:thr2}
n_S \leq \Big(\frac{\log(\tau_1/\tau_0)}{\log((1-\tau_0)/(1-\tau_1))}+1\Big)^{-1} n_I\doteq n_{th}(n_I) .
\end{equation}
Using Eq. (\ref{ps}), the probability of success 
for the TMSV-based transmitter and 
photon-counting receiver is:  
\begin{equation}
P_s^{TMSV,pc}(\tau_0,\tau_1)=\frac{1}{2}\sum_{n_I=0}^\infty\sum_{n_S=0}^{n_{th}(n_I)} P(\textbf{n}|\tau_0) + \frac{1}{2}\sum_{n_I=0}^\infty\sum_{n_S=n_{th}(n_I)}^\infty P(\textbf{n}|\tau_1).
\end{equation}

Remarkably, the numerical simulations of Ref.~\cite{ortolano20} demonstrate that this approach offers a gain respect to the optimal classical bound in Eq. (\ref{cl}) and the classical photon counting bound in Eq.~(\ref{eq:coh}). Note that the numerical investigation shows a quantum advantage even with a single TMSV
state but the discrimination is
more effective in presence of an high number of copies $M \gg 1$. This is the regime that we have in our experiment. 
Note that when the number of modes $M$ is  much higher than the mean number of photons $n$ and at the same time $n\gg1$, the photon number distributions are well approximated by a multivariate normal distribution. We will use this approximation in our analysis

\subsection{Effect of quantum efficiency}
\label{sec:eta}
A quantity characterizing every quantum optical experimental setup is the quantum efficiency $0 \leq \eta \leq 1$. This quantity expresses the fraction of photons entering the experimental setup that are actually detected. Being interested in the quantum reading experimental realisation this quantity is of particular interest. It takes into account photon losses from interaction with the environment and optical components as well as the intrinsic quantum efficiency of the detector used and  its electronic noise \GO{and, in case of correlated beams, the efficiency in detecting correlated photons can be accounted by this coefficient as well}. 

It is well known that quantum properties require a good quantum efficiency $\eta$ of the setup to be revealed and used to get an advantage over classical limits, and that those advantages are usually very sensitive to variations in the value of $\eta$. An analysis of its effect is then necessary to obtain accurate predictions of the experimental outcome. 

The imperfect experimental setup acts as an amplitude attenuation channel, with transmission $\eta$, on the input quantum state. The effect of an imperfect quantum efficiency is therefore indistinguishable from the attenuation coming from the memory cell. From the point of view of the photon number distribution this is stated by the composition property of the binomial distribution characterizing the process:
\begin{align}
\sum_{m=n}^N B(m|N,\tau)B(n|m,\eta)&= \sum_{m=n}^N \genfrac(){0pt}{0}{N}{m} \genfrac(){0pt}{0}{m}{n}  \tau^m(1-\tau)^{N-m}\eta^n(1-\eta)^{m-n}=\nonumber \\
&=\genfrac(){0pt}{0}{n}{N}\sum_{m=n}^{N}     \genfrac(){0pt}{0}{N-n}{m-n} \tau^m(1-\tau)^{N-m}\eta^n(1-\eta)^{m-n} = \nonumber \\
&=\genfrac(){0pt}{0}{n}{N}(\tau\eta)^n(1-\tau)^{N-n}\sum_{\alpha=0}^{N-n}  \genfrac(){0pt}{0}{N-n}{\alpha} \Big(\frac{\tau(1-\eta)}{1-\tau}\Big)^{\alpha}=  \nonumber \\
&=\genfrac(){0pt}{0}{n}{N}(\tau\eta)^n(1-\tau)^{N-n}\Big(1+\frac{\tau(1-\eta)}{1-\tau}\Big)^{N-n}=  \nonumber \\
&=\genfrac(){0pt}{0}{n}{N}(\tau\eta)^n(1-\tau\eta)^{N-n}=  \nonumber \\
&=\sum_{m=n}^N B(m|N,\eta)B(n|m,\tau)=B(n|N,\tau\eta).
\end{align}
In view of this property a generalization of the result of the previous section to a \emph{signal system} with an efficiency $\eta_s<1$ is readily found by performing the substitution:
\begin{equation}
\tau_\alpha \rightarrow \eta_s\tau_\alpha.
\end{equation}
As said above, this is due to the indistinguishability of the two attenuation processes, $\eta_s$ and $\tau_\alpha$, that the quantum state undergoes; it shows that the discrimination of two quantum channels $\tau_0$ and $\tau_1$ with a signal quantum efficiency $\eta_s$ is equivalent to the discrimination between two channels $\eta_s\tau_0$ and $\eta_s\tau_1$ with perfect quantum efficiency. This argument does not depend on the measurement performed, so it is valid also in the case of the classical theoretical bound in Eq.~(\ref{cl}), i.e. when computing it the effective channels to be considered are $\eta_s\tau_0$ and $\eta_s\tau_1$.

\section{EXPERIMENTAL REALIZATION}
\label{sec:sections}
 
\subsection{Experimental set-up}
\begin{figure} [ht]
\centering
\includegraphics[width=0.7\textwidth]{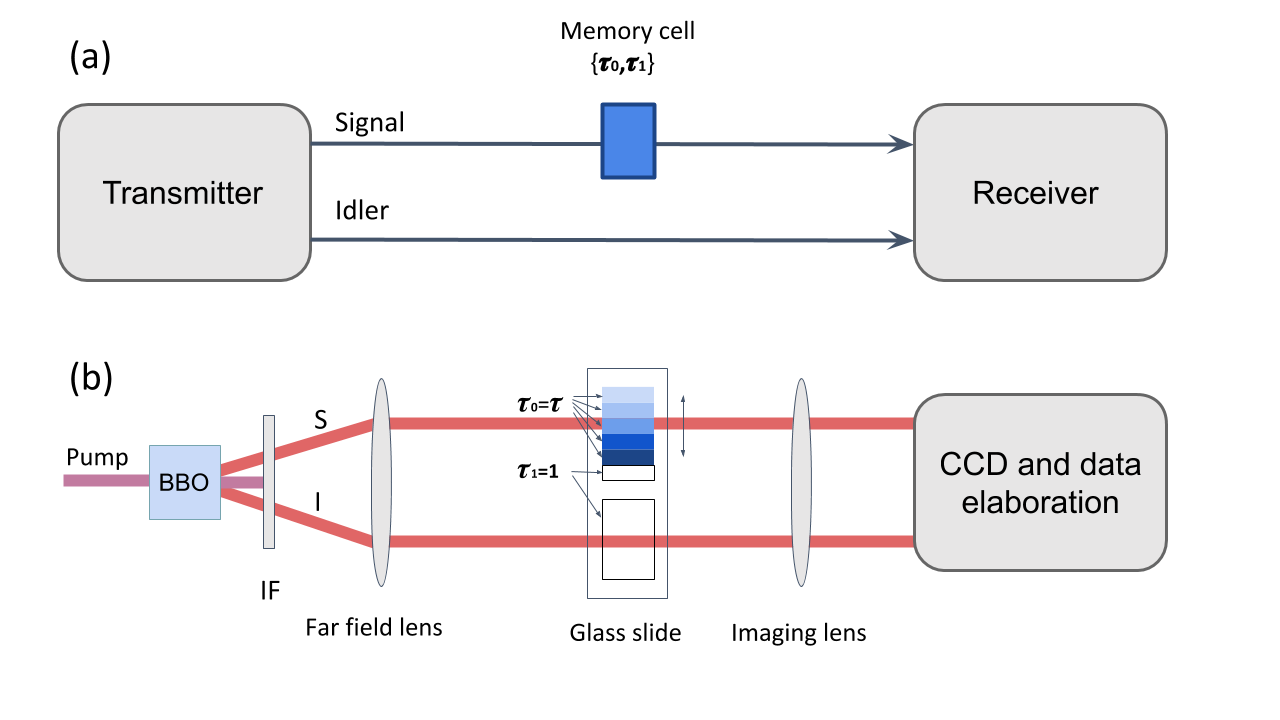}
\caption{\label{fig:setup}(a) Model of the reading of a classical digital memory (b) Simplified schematic of the experimental setup. In the BBO crystal the multi-mode squeezed vacuum state is generated. The probe (signal) beam passes through the memory cell realized by an absorbing layer with two value of transmittance, $\tau_0$ or $\tau_1$ and is addressed to a region of the the CCD camera chip. The reference (idler) beam goes directly to an other region the chip, without interacting with the cell. The integrated signals from the two regions, $n_S$ and $N_R$ respectively, are recorded in each run. BBO: Type-II-Beta-Barium-Borate non linear crystal. IF: interferential filter ($800 \pm 20$)nm. CCD: charge-coupled device camera. }
\end{figure}

A scheme of the experimental set-up is reported in Fig. \ref{fig:setup}(b). Comparing it with Fig. \ref{fig:setup}(a) it can be appreciated how the different elements involved in the quantum reading model are experimentally implemented. 

We experimentally produce a multi-mode squeezed vacuum state of light exploiting the spontaneous parametric down conversion process in a non linear crystal. More specifically, we pump a $(1\mathrm{cm})^3$ type-II-Beta-Barium-Borate crystal with a CW laser of $\lambda_p=405$nm and power of $100$mW. The laser emission is triggered by a digital signal coming from the detector, i.e. the laser emits only in correspondence of the acquisition time. An interferential filter at $(800 \pm 20)$nm performs a spectral selection of the down-converted photons allowing only the photons around the degenerate frequency ($\lambda_d=2 \lambda_p=810$nm) to reach the detector. Using a system of two lenses, the correlation in momentum of two down-converted photons is mapped into spatial correlations at the detection plane. In particular, the far-field plane of the emission, where spatial correlation occurs, is firstly realized in the focal plane of the far-field first lens, having focal length $f_{FF}=1$cm. Using a second system of lenses with equivalent focal length of about $f_{IM}=1.6$cm, named as imaging lens in the figure, the far-field plane is imaged to the detection plane, with a magnification factor of $M=7.8$. The detector is a charge-coupled-device (CCD) camera (Princeton Instrument Pixis 400BR Excelon), working in linear mode, with high quantum efficiency (nominally $> 95\%$ at $810$nm), $100\%$ fill factor and low electronic noise. We work at the 100kHz digitization rate, to keep the noise as low as possible. The physical pixels of the camera measure $13 \mu$m, nevertheless in this experiment, where spatial resolution is not crucial, we group them by $12 \times 12$ hardware binning, thus lowering the acquisition time and the read-out noise. 

\begin{figure} [ht]
\centering
\includegraphics[width=0.7\textwidth]{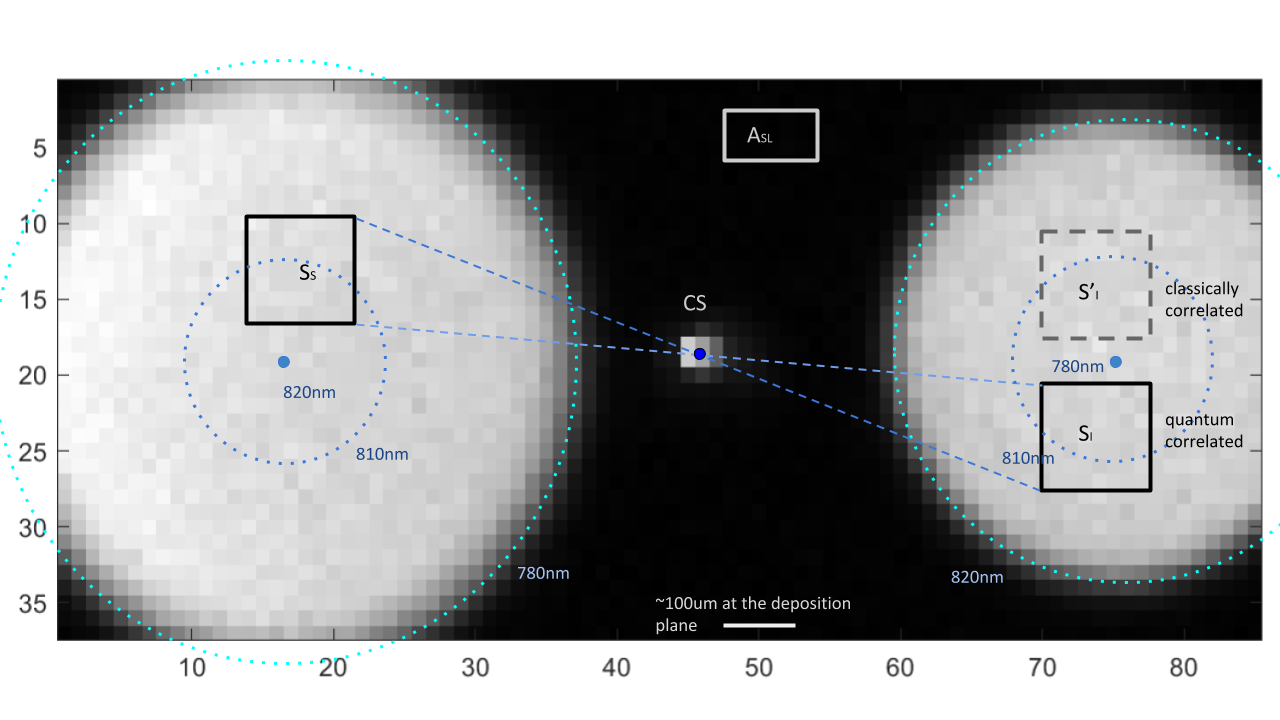}
\caption{\label{fig:ccd}Typical frame acquired at the CCD camera. $S_S$ and $S_I$ indicate the two detection areas considered: being symmetric respect to the center of symmetry (CS), they present quantum photon number correlation. On the contrary, $S_S$ and $S'_I$ are only classically correlated. The average number of counts per pixel in these regions is of the order of $10^4$, integrating on the areas considered the values $n_S$ and $n_I$ are obtained and used in subsequent elaboration. $A_{SL}$ is the region used to evaluate the straylight. The different dotted cicles indicate the different wavelength of the photons in their correspondence. Due to the interferential filter used, photon of $780 \mathrm{nm} \leq \lambda \leq 810 \mathrm{nm}$ are detected. The degeneracy frequency is $\lambda_d=2\lambda_{pump}=810 \mathrm{nm}$.}  
\end{figure}

In Fig. \ref{fig:ccd} a typical frame acquired by the CCD is reported: $S_S$ and $S_I$, symmetric respect to the center of symmetry, indicate the two detection areas considered, corresponding to the signal and idler beams respectively. In our case these areas contain $\sim 50$ macro-pixels and collect $M_{sp}=8 \cdot 10^{2}$ spatial modes, where the linear size of a mode has been evaluated by an intensity spatial correlation measurement to be of about 5$\mu$m. For the further data elaboration we integrate the signals in $S_S$ and $S_I$ and we name as $n_S$ and $n_I$ the number of photons detected in the two regions respectively. 
To estimate the number of temporal modes collected in a single frame we can compare its acquisition time, $\sim 20$ms, with the coherence time of the SPDC process, $10^{-12}$s. It results $M_{t}=2 \cdot 10^{10}$. The favourable condition of working with an high number of modes $M=M_{sp} \cdot M_t \gg 1$ is thus satisfied. Considering that we typically work with $\langle n_S \rangle \sim 3 \cdot 10^5$, we can conclude that the number of photons per spatio-temporal mode is $\nu = \frac{\langle n_S \rangle}{M_{sp} \cdot M_t}  \sim 2 \cdot 10^{-8} \ll 1$.

To reproduce the quantum-reading scheme presented in Fig. \ref{fig:setup}(a) we insert in the focal plane of the first lens a coated glass-slide, playing the role of the digital memory support. It presents a deposition of variable transmission $\tau_0$ ($0.990 < \tau_0 < 1$), while outside the deposition the transmission is $\tau_1 =1$. The probe beam interception with the glass slide represents the single cell memory. Its transmittance can assume two values, $\tau_0$ or $\tau_1$, depending on the presence or absence of the deposition respectively, and therefore can be used to store a bit of information. In order to have the same optical path of the probe beam, also the reference beam passes through the glass, but never intercepts the deposition.

\begin{figure} [ht]
\centering
\includegraphics[width=0.7\textwidth]{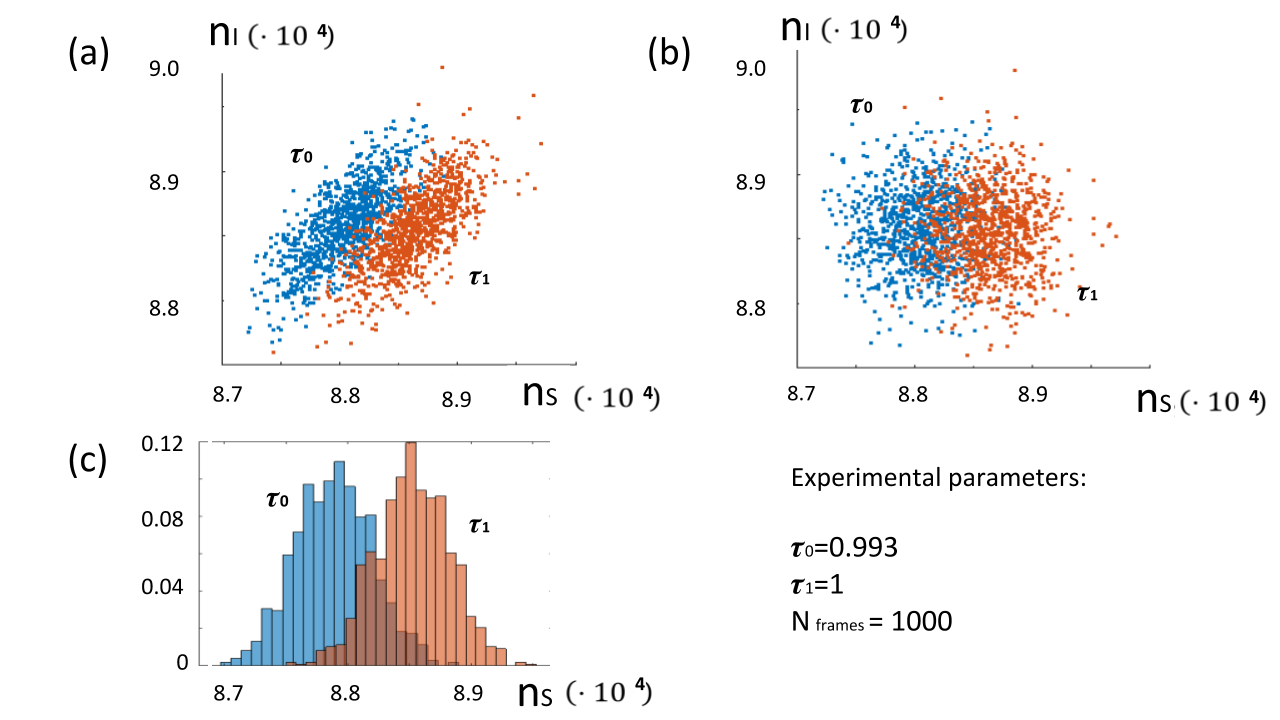}
\vspace{+0.4cm}
\caption{\label{fig:twb_corr}(a) $n_I$ in function of $n_S$, for 1000 frames. In this case the regions $S_I$ and $S_S$, symmetric respect to the center of symmetry, are considered. Blue data correspond to $\tau_0 \sim 0.993$, while red data corresponds to $\tau_1=1$. (b) $n'_I$ in function of $n_S$, for 1000 frames. In this case the regions $S'_I$ and $S_S$ are considered. Blue dots correspond to $\tau_0 \sim 0.993$, while red dots corresponds to $\tau_1=1$.(c) $n_S$ relative frequency distribution for $\tau_0 \sim 0.993$ (blue histogram) and $\tau_1=1$ (red histogram).}
\end{figure}

The two beams of a two mode squeezed vacuum state present non-classical photon number correlation given by the entangled nature of the state. This is demonstrated from the squeezed shape of the joint distributions in Fig. \ref{fig:twb_corr} (a), where it is reported $n_I$ as function of $n_S$, for 1000 frames. Blue data correspond to $\tau_0=0.993$  while red data correspond to $\tau_1=1$. For comparison, Fig. \ref{fig:twb_corr} (b) shows $n_I$ as function of $n_S$, for 1000 frames, in absence of quantum correlation (e.g. what we could obtain splitting a classical beam with a 50\% beam splitter). Since both signal and idler beams are already shot noise limited from the practical viewpoint (more precisely they follow a multi-thermal statistics with a number of photons per mode $\nu\ll1$, which approaches a Poisson's distribution), further squeezing of the distribution does not occur if quantum correlation are not present. To experimentally reproduce this scenario with our set-up we consider two regions on the detector, $S_S$ and $S'_I$, not symmetric respect the center of symmetry (see the dashed square in Fig. \ref{fig:ccd}). 

In Fig. \ref{fig:twb_corr} (c) it is reported the frequency histogram of $n_S$, for $\tau_0=0.993$ (blue data) and $\tau_1=1$ (red data). This is the information we have access if we consider only the signal beam, which alone does not present any quantum feature. While in Fig. \ref{fig:twb_corr} (a) the two distributions, for $\tau_0$ and $\tau_1$ respectively, are well separated, in Fig. \ref{fig:twb_corr} (c) they significantly overlap. Intuitively, this difference is at the basis of the quantum enhancement offered by using the TMSV state rather than a single classical beam.

We implement the readout memory strategy exploiting two mode squeezed vacuum intensity correlations presented in Sec. \ref{sec:twb} and we experimentally evaluate its information gain respect to both the classical bound in Eq. (\ref{cl}), $G_a$, and the classical single beam photon-counting strategy discussed in Sec. \ref{sec:cl} ($G_{emp}$). Different number of photons $N$ probing the memory cell and different values of $\tau_0$ are considered. To change $\tau_0$ we simply consider depositions of different absorption by moving the glass slide appropriately. For changing the number of probing photons we simply change the dimension of the $S_S$ and $S_I$ regions, avoiding the necessity of acquiring new data with a different laser power. 
In a calibration phase, we estimate all the parameters necessary in the subsequent analysis: $N$, $\tau_0$, $\eta_S$, $\eta_I$, being $\eta_S$ and $\eta_I$ the channel efficiency on the signal and idler beam respectively. The method used to estimate these efficiencies exploits the peculiarity of the two-mode squeezed vacuum state and is discussed in Sec. \ref{sec:etaexp}. The values of the experimental parameters estimated in the calibration phase are used to compute the theoretical classical bounds and to compute the probabilities $P_{\eta_{S},\eta_{I}}(\textbf{n}|\tau)$  necessary for applying the Bayesian decision strategy presented in Sec. \ref{bd}

For the experimental estimation the error probability $P_{err}$ while using the two mode squeezed vacuum state we proceed as follows: we consider two sets of frames (in our case 10000 frames per set are acquired), one for $\tau=\tau_0$ and one for $\tau=\tau_1$. For each frame, we estimate the value of the memory cell according to the measured ($n_S,n_I$) and  the strategy discussed in Sec. \ref{sec:twb}. Repeating this procedure for all the frames in both sets and considering the ratio between the number of incorrect choices and the total number of frames considered, we estimate the error probability $P_{err}$. A similar approach can be adopted for evaluating the error probability in the single beam classical strategy, simply considering only $n_S$ and the strategy presented in Sec. \ref{sec:cl}. 

From $P_{err}$ we can straightforward calculate the associated binary entropy $H(P_{err})$ and the information gain of the two mode squeezed vacuum strategy over both the optimal classical bound ($G_a$, see Eq. (\ref{eq:gain})) and the single-beam classical approach ($G_{emp}$, see Eq. (\ref{eq:gemp})). 

\begin{figure}
     \centering
     \begin{subfigure}[b]{0.44\textwidth}
         \centering
         \includegraphics[width=\textwidth]{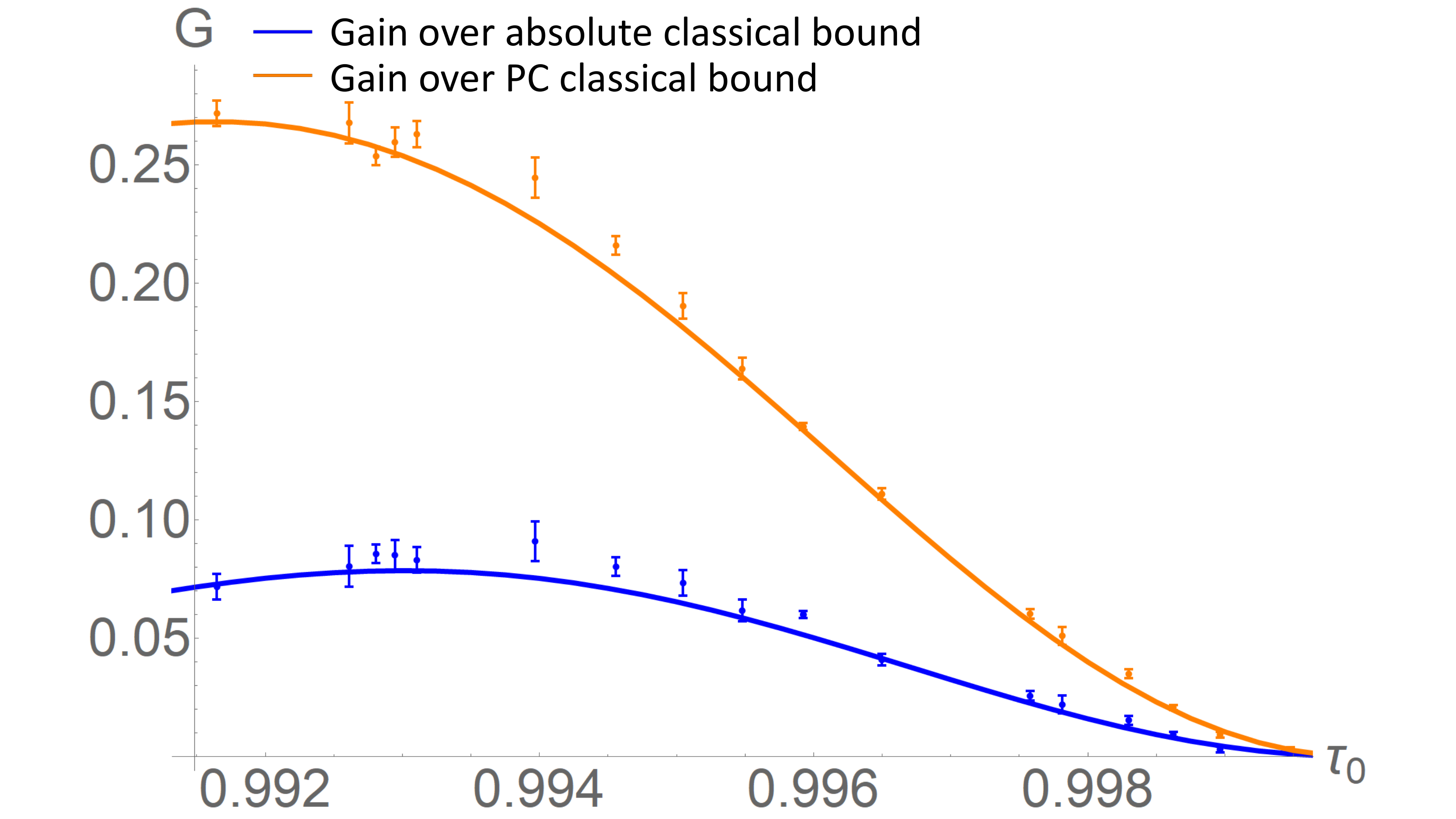}
         \caption{$N=1.15 \cdot 10^5$}
         \label{fig:y equals x}
     \end{subfigure}
     \hfill
     \begin{subfigure}[b]{0.44\textwidth}
         \centering
         \includegraphics[width=\textwidth]{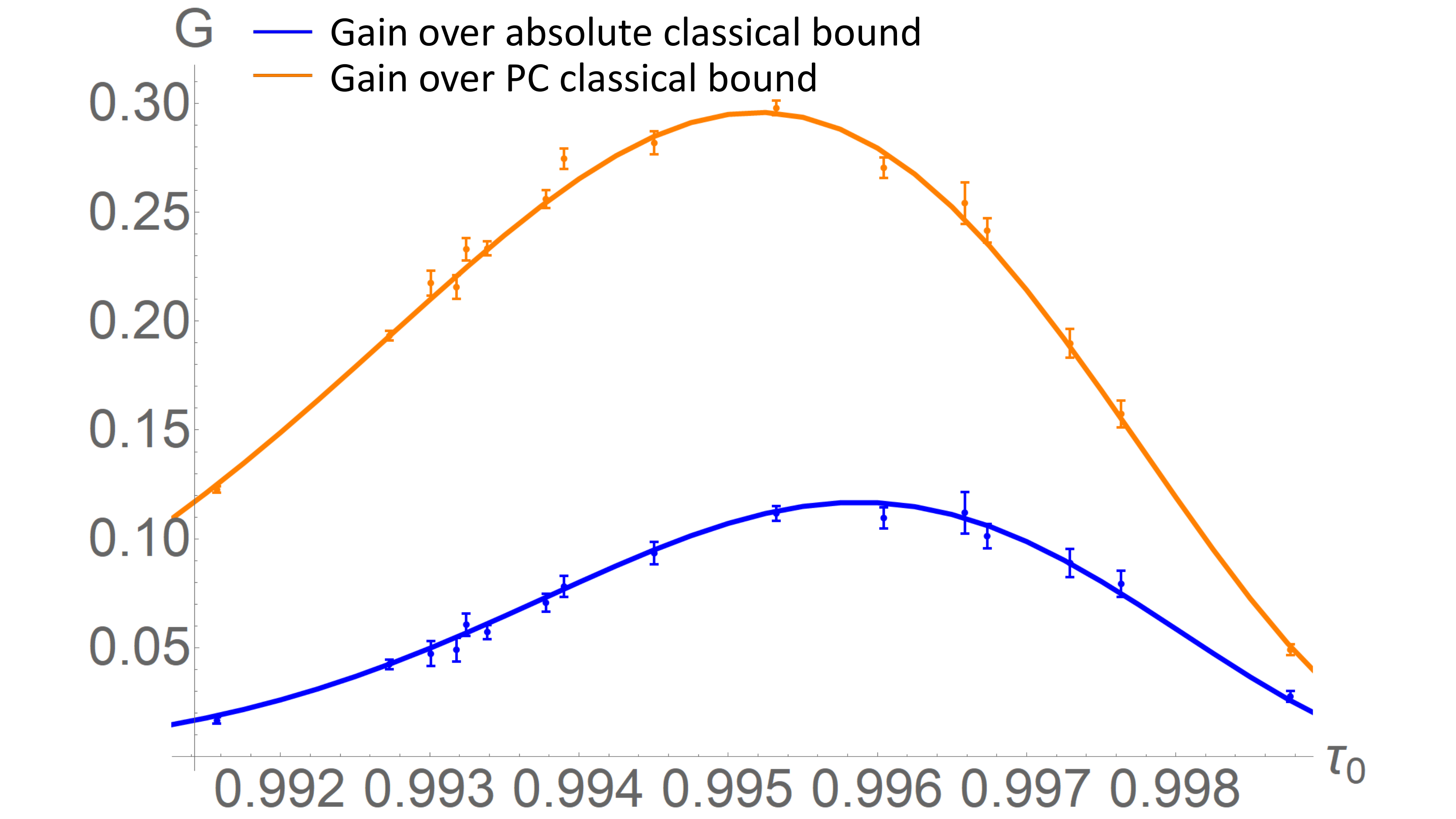}
         \caption{$N=3.1 \cdot 10^5$}
         \label{fig:three sin x}
     \end{subfigure}
     \hfill
     \begin{subfigure}[b]{0.44\textwidth}
         \centering
         \includegraphics[width=\textwidth]{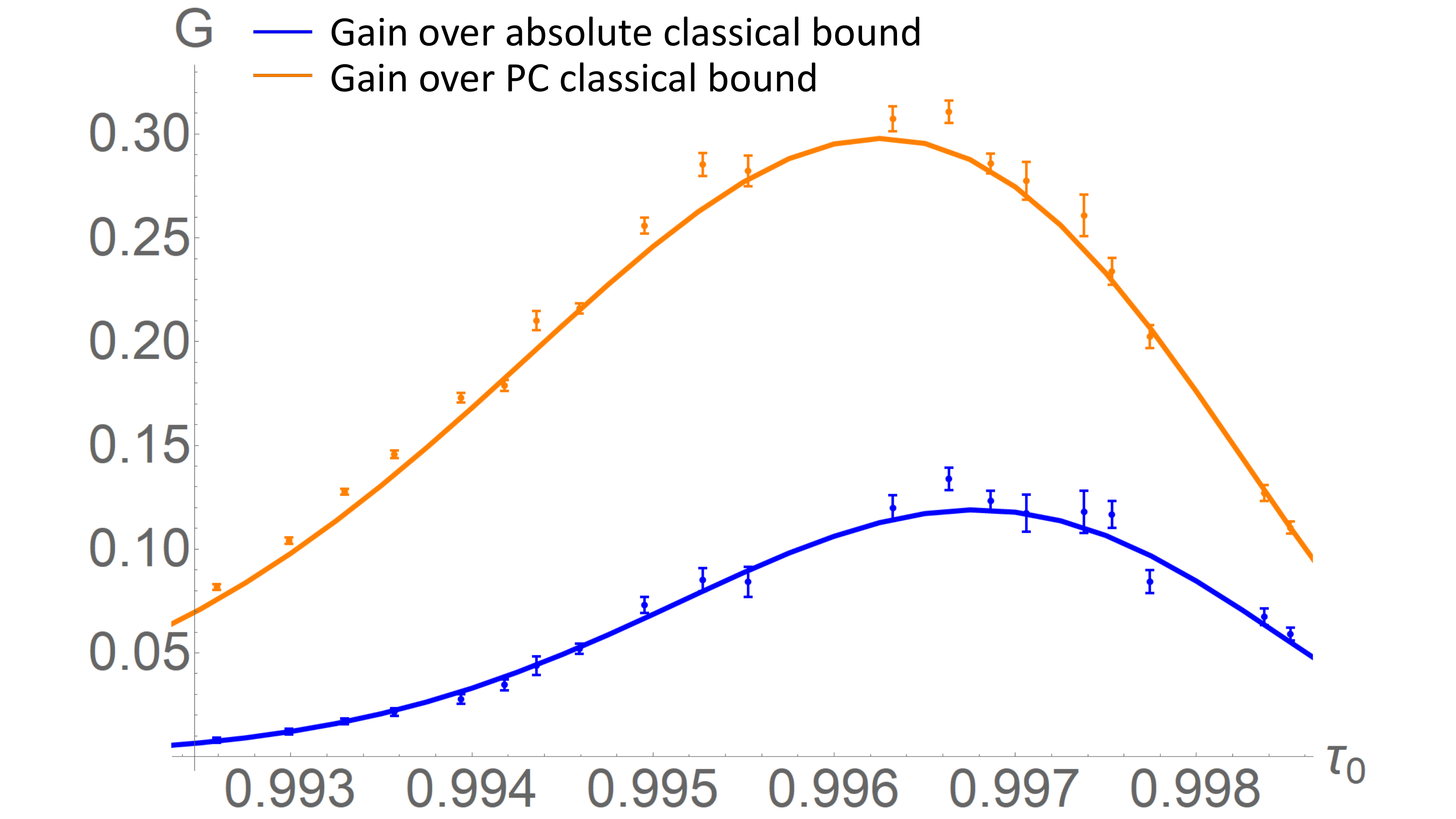}
         \caption{$N=5.2 \cdot 10^5$}
         \label{fig:five over x}
     \end{subfigure}
     \vspace{+0.4cm}
        \caption{\label{fig:exp_res}Experimental gain $G$ of quantum reading (bits) as a function of the lower transmissivity $\tau_0$. The three panels are obtained for different mean photon number in the signal beam: (a) $N=1.15 \cdot 10^5$, (b) $3.1 \cdot 10^{5}$, and (c) $5.2 \cdot 10^{5}$. Blue data refers to the gain with respect to the classical optimal bound in Eq.~(\ref{cl}). Red data refers to the gain with respect to the classical photon-counting bound given in Eq.~(\ref{eq:coh}), obtained from the marginal distribution of the signal. The error bars represent the experimental uncertainty in the estimated gains at $1\sigma$. The mean detection efficiency of signal and idler channels $\eta_S$ and $\eta_I$ are $\eta_S=0.78$ and $\eta_I=0.77$. The electronic noise is estimated to be $\nu_e\sim10^4$.}
\end{figure}

The experimental results are reported in Fig. \ref{fig:exp_res}.


The three figures are obtained for a different number of photons in the signal beam: $\langle N \rangle \sim 1.15 \cdot 10^{5},3.1 \cdot 10^{5}$ and $5.2 \cdot 10^{5}$ respectively. Blue data represents $G_a$, while red data $G_{emp}$. The error bars at $1\sigma$ are obtained dividing the  data into 10 sets and independently evaluating the gains in each set. The theoretical curves obtained using the theoretical model discussed above are also reported. 

We can conclude that a quantum advantage in the regime under consideration is present. In particular, we experimentally demonstrate a maximum quantum gain of larger than 0.1 bit per cell over the classical optimal bound, and of $\sim 0.3$ over the classical photon counting strategy. These results paves the way for future experimental applications.


\subsubsection{Efficiency estimation}\label{sec:etaexp}
Among other quantities, the estimation of the quantum efficiency is particularly crucial.
To estimate it, we use an extension of the Klishko method \cite{klyshko80}, extensively discussed in Ref.~\cite{brida10}. The key idea of this method is that the detected photon number correlation between signal and idler while considering a two-mode squuezed vacuum state is not perfect because $\eta$ is not unitary: from the degradation of the the photon number correlation we can infer the efficiency of the sistem. In the following we summarize the operative procedure used to estimate $\eta_S$ and $\eta_I$, i.e. the quantum efficiency for the signal and idler modes respectively.

Said $S_S$ and $S_I$ two conjugated regions at the detector, in the idler and signal beam respectevely, 
we introduce the quantity:
\begin{equation}
    \gamma = \frac{\langle n_S \rangle}{\langle n_I \rangle} = \frac{\eta_S}{\eta_I},
\end{equation}
where $\langle n_S \rangle$ and $\langle n_I \rangle$ are the mean number of photons detected in $S_S$ and $S_I$ respectively. This quantity, which represents the unbalancement between the two channels, is proportional to the ratio of the two efficiencies. $\gamma$ can be easily experimentally estimated and allows balancing the photon counts of the two channels.
In addition to losses, there are two main other sources of noise:
\begin{itemize}
    \item Stray-light, $\langle N_{SL} \rangle$: it is due to fluorescence of the laser pump in the BBO crystal and in the interference filter. It can be estimated considering the mean number of photons detected per pixel in a "dark" region, as $A_{SL}$ in Fig. \ref{fig:ccd}.
    \item Electronic noise, $\Delta^2_{el}$: it can be estimated from frames acquired with the shutter closed, where the only noise source is the electronic noise. 
\end{itemize}
We can now define $\sigma$ which quantifies the degree of correlation between signal and idler, taking into account the efficiency unbalancement and the presence of noise:
\begin{equation}
    \sigma_{\gamma,B}=\frac{ \Delta^2(n_S - \gamma n_I) }{\langle n_S + \gamma n_I \rangle} \cdot \frac{\langle n_S \rangle}{\langle n_S - N_{SL} \rangle} - \frac{\Delta^2_{el} + \langle N_{SL} \rangle}{\langle n_S - N_{SL} \rangle}.
\end{equation}

It turns out that, considering $S_S$ and $S_I$ sufficiently big respect to the coherence area of the spontaneus parametric down conversion process, $\sigma_{\gamma,B}$ can be written in terms of $\eta_S$ as:
\begin{equation}\label{eq:sigmaeta}
    \sigma_{\gamma,B}=\frac{1+\gamma}{2}-\eta_S .
\end{equation}
Inverting the expression in Eq. (\ref{eq:sigmaeta}) and using the definition of $\gamma$, the two efficiencies can thus be estimated as:
\begin{equation}
    \eta_S = \frac{1+\gamma}{2} - \sigma_{\gamma,B},
\end{equation}
\begin{equation}
    \eta_I=\frac{\eta_S}{\gamma}.
\end{equation}
The uncertainty on this estimate can be experimentally evaluated repeating the measurement several times and then considering the standard deviation on the values obtained.

\acknowledgments 
 
This work has been sponsored by the EU Horizon 2020 FET-Open project: “Quantum readout techniques and technologies” (QUARTET, Grant agreement No 862644). 

\bibliography{report} 
\bibliographystyle{spiebib} 

\end{document}